# A Sampling-Based Approach to Computing Equilibria in Succinct Extensive-Form Games

**Miroslav Dudík, Geoffrey J. Gordon**
Carnegie Mellon University
Machine Learning Department
5000 Forbes Avenue, Pittsburgh, PA 15213
{mdudik,ggordon}@cs.cmu.edu

## Abstract

A central task of artificial intelligence is the design of artificial agents that act towards specified goals in partially observed environments. Since such environments frequently include interaction over time with other agents with their own goals, reasoning about such interaction relies on sequential game-theoretic models such as extensive-form games or some of their succinct representations such as multi-agent influence diagrams. The current algorithms for calculating equilibria either work with inefficient representations, possibly doubly exponential in the number of time steps, or place strong assumptions on the game structure. In this paper, we propose a sampling-based approach, which calculates extensive-form correlated equilibria with small representations without placing such strong assumptions. Thus, it is practical in situations where the previous approaches would fail. In addition, our algorithm allows control over characteristics of the target equilibrium, e.g., we can ask for an equilibrium with high social welfare. Our approach is based on a multiplicative-weight update algorithm analogous to AdaBoost, and Markov chain Monte Carlo sampling. We prove convergence guarantees and explore the utility of our approach on several moderately sized multi-player games.

## 1 INTRODUCTION

The goal of artificial intelligence is the design of artificial agents achieving specified objectives in real-world environments. Such environments often include interactions with other agents (artificial or human) with their own goals. Game theory provides a useful framework to reason about such interactions. Unfortunately, the majority of current computational techniques were developed for one-step games, in which each agent takes a single action without observing actions of others. Such techniques are inadequate in more realistic settings when the interaction happens over time and agents can partially observe actions of others, such as in a game of poker, in a political negotiation, or in a business interaction among multiple parties. These settings are traditionally modeled as extensive-form games (EFGs) [Kuhn, 1953], described in detail in Section 2.

Techniques developed for one-step games suffer from two principal drawbacks when applied to EFGs. First, their representations are linear in the number of possible deterministic behaviors of a single agent across all possible situations that can arise during the game. Since the number of possible situations can grow exponentially in the number of time steps, the number of possible deterministic behaviors can grow doubly exponentially in the number of time steps. We seek to limit this explosion. Second, solution concepts which are tractable and desirable in one-step games (such as correlated equilibria) become intractable or undesirable in sequential settings. We seek to develop tractable solutions since, concurring with Papadimitriou [2005], we believe that "[i]ntractability of an equilibrium concept would make it implausible as a model of behavior."

Several existing approaches partially address these challenges. The *sequence-form representation* [Koller and Meggido, 1992] is linear in the number of possible situations rather than behaviors, but in order to create this representation without unfolding the entire game-tree, the underlying EFG is typically required to have a special structure, for example, be represented as a multi-agent influence diagram [Koller and Milch, 2001]. The *extensive-form correlated equilibrium* [Forges and von Stengel, 2002] is a solution concept both more tractable and more natural in sequential games than concepts developed for one-step games. However, existing algorithms for computing this type of equilibrium either only treat special cases, such as two-player games without chance moves [Forges and von Stengel, 2002], or delegate potentially intractable subproblems to an outside oracle [Gordon et al., 2008].

In this paper, we propose a sampling-based approximate approach to finding extensive-form correlated equilibria. Our approach is fully general in that it applies to multi-player games with chance moves and arbitrary utility functions. Our technique is applicable to very large EFGs without imposing any requirements on their structure, except for the standard assumption of perfect recall and the ability to "simulate the game", i.e., the ability to determine how an action in a given situation leads to a new situation. We formalize this notion by defining *succinct EFGs* in Section 2.1. The space complexity of our algorithm is no worse than the space complexity of algorithms developed for special types of EFGs, and in some cases our representations may be significantly smaller. Thus, we get the benefits of the sequence-form representation without imposing strong assumptions on the underlying game.

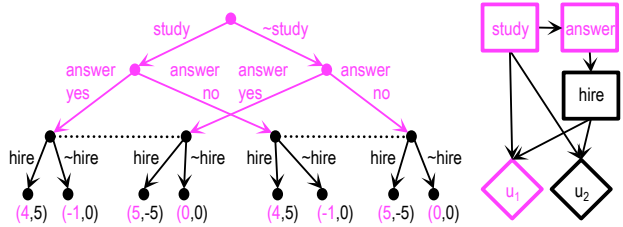

Figure 1: *The job market game. Left:* An EFG representation. Nodes connected by dotted lines belong to the same information sets. *Right:* A MAID representation. Decision nodes are depicted as rectangles, utility nodes as diamonds.

Our algorithm finds extensive-form correlated equilibria of maximum entropy or minimum relative entropy, possibly under additional constraints. This formulation yields boosting-style updates with a favorable convergence rate. In addition, the possibility of including additional constraints allows finding equilibria of high social welfare or equilibria that are consistent with an observed behavior. As subroutines, our algorithm uses best-response calculations and MCMC sampling. We explore the utility of our algorithm on three examples: a two-player toy example, a game with a large number of weakly interacting players, and a three-player poker variant. Our poker variant uses a deck of eight cards and includes one round of betting. This is quite small compared with state of the art for two-person zero-sum games (see, e.g., [Gilpin and Sandholm, 2006]). However, nonzero-sum or multi-player games are generally believed to be much more challenging than zero-sum games, and we believe that no existing technique would be able to calculate equilibria for our example.

## 2 DEFINITIONS

Let $N$ be the number of players, denoted as $n = 1, \ldots, N$; randomness is modeled as an additional "nature" player, denoted *nat*; when we say "player" we typically mean a regular player $n$ and refer to nature explicitly. An *extensive-form game* (EFG) is represented by a *game tree*, where inner nodes are partitioned into *information sets*. Each information set $i$ belongs to a unique player $n$ or nature, who is required to act upon reaching nodes in that information set. The player knows only the identity of the information set, but cannot distinguish among the nodes in it; thus, information sets represent partial information. Nodes in the same information set have the same number of outgoing edges, corresponding to actions that a player can take. The game begins in the root of the tree; players and nature take turns until reaching a leaf. Each leaf $\ell$ contains an assignment of utilities $u_n(\ell) \in \mathbb{R}$ to individual players, jointly denoted $\mathbf{u}(\ell) \in \mathbb{R}^N$. Collections of information sets of players and nature are denoted $I(n)$ and $I(nat)$. The number of actions available in the information set $i$ is $A_i$. We assume that $A_i \geq 2$ and denote actions as $a = 1, \ldots, A_i$.

A deterministic behavior of the player $n$ is described by a *pure strategy*, which is a tuple $s_n = (s_i)_{i \in I(n)}$ with $s_i \in \{1, \ldots, A_i\}$ specifying which action to take in each information set. The vector $s = (s_n)_{n \leq N}$ of pure strategies of all players is referred to as the *strategy profile*. Nature's pure strategies, denoted $s_{nat} = s_{I(nat)}$ (we use set subscripts to denote tuples), are referred to as *scenarios*. We assume that scenarios are sampled from a fixed distribution $p_{nat}$ which factors as $p_{nat}(s_{nat}) = \prod_{i \in I(nat)} p_i(s_i)$. As common in game theory, we restrict our attention to EFGs with *perfect recall*. Perfect recall means that players do not forget any information over the course of the game. Formally, this requires that paths reaching nodes in an information set $i \in I(n)$ are indistinguishable by $n$, i.e., they contain identical sequences of information sets of the player $n$, and the player $n$ took identical actions in those information sets. For nature's information sets we only assume that the same information set does not appear on the same path more than once (this is a weaker condition than perfect recall).

Given a strategy profile $s$ and a scenario $s_{nat}$, the game tree reaches a unique leaf $\ell$; we write $\mathbf{u}(s, s_{nat})$ for the utility vector $\mathbf{u}(\ell)$ achieved at $\ell$. Taking expectation over nature's randomness, the vector of expected utilities is defined as

$$\mathbf{u}(s) = \mathbb{E}_{nat}[\mathbf{u}(s, s_{nat})] \ .$$

The function $\mathbf{u}(s)$, defined implicitly by an EFG, is referred to as the *strategic form* of the EFG.

Some components of a pure strategy $s_n$ for the player $n$ may be irrelevant—the components specifying actions for information sets that cannot be reached given the actions of $n$ earlier in the game. By setting such components to a "don't care" value $*$, we obtain a smaller set of deterministic strategies called *reduced strategies*, and denoted $s_n^*$.

Let $S_n$ and $S_n^*$ denote sets of pure and reduced strategies of $n$, and $S$ and $S^*$ sets of pure and reduced strategy profiles, respectively. Algorithms developed for strategic-form games (i.e., games where each player takes only one action and cannot observe actions of others) typically require time and space polynomial in $|S^*| = \prod_n |S_n^*|$, or, in some cases (e.g., [Papadimitriou, 2005]), $\sum_n |S_n^*|$, which may still be prohibitively large. A more efficient class of algorithms developed specifically for EFGs uses the *sequence-form representation*, which is equivalent to the randomized *behavioral strategy*, describing for each information set of a given player the probability of taking an action $a$. We refer to the size of this representation as the *sequence complexity*, denoted $\Gamma = \sum_n \Gamma_n$, where $\Gamma_n = \sum_{i \in I(n)} A_i$. Note that $|S_n| = \prod_{i \in I(n)} A_i$, hence we obtain

$$|S_n^*| \leq |S_n| < 2^{\Gamma_n} \quad , \quad |S^*| \leq |S| < 2^{\Gamma}$$

(we used $A_i < 2^{A_i}$ for $A_i \in \mathbb{N}$). Since $A_i \geq 2$, we also obtain that $\Gamma_n/2 \leq |S_n^*|$. It is not too difficult to construct examples where $|S_n^*| \geq 2^{\Omega(\Gamma_n)}$ [von Stengel et al., 2002]. Hence, algorithms based on sequence form are usually significantly more efficient than algorithms working with strategic form.

Fig. 1 gives an example of an EFG corresponding to the *job market game* between the *student* and the *employer* (discussed by Spence [1973] and others, our version along the lines of Forges and von Stengel [2002]). In this game, the student first decides to study or not to study. The student then comes to a job interview and is asked a question. The employer hears the student's answer and decides whether to hire the student or not. The student benefits by being hired and suffers slightly by studying. The employer benefits by hiring a student that studied and suffers by hiring one who did not. We use this game as a running example, but our goal is to reason about much larger EFGs.

### 2.1 SUCCINCT EXTENSIVE-FORM GAMES

We study EFGs whose game trees are potentially too large to be stored in memory explicitly. Therefore, we work with implicit representations which we call *succinct EFGs*.

We define the *type* of an EFG as a pair $(\Gamma, r_{\max})$ where $\Gamma$ is the sequence complexity (or any upper bound on sequence complexity) and $r_{\max} \in \mathbb{R}$ is a *regret bound*. We define regret formally below, but here we note that it is bounded by the range of a player's utility values. By a *succinct EFG*, we mean a representation of an EFG which supports the following queries:

- $(\Gamma, r_{\max})$, the type,
- $h_{start}$, the root of the game tree,
- $infoset(h)$, the information set corresponding to the node $h$; value $nil$ is returned when $h$ is a leaf,
- $player(i)$, the player (or nature) acting in the information set $i$,
- $A_i$, the number of actions in the information set $i$,
- $next(h, a)$, the node reached after taking action $a$ in $h$,
- $p_i(a)$, nature's randomness (only defined if $player(i) = nat$),
- $\mathbf{u}(\ell)$, the utility vector in the leaf $\ell$.

A prime example of succinct EFGs are *multiagent influence diagrams* (MAIDs) [Koller and Milch, 2001]. Multiagent influence diagrams are game-theoretic generalizations of Bayes nets. Similar to Bayes nets, MAIDs are represented by directed acyclic graphs. They have three types of nodes: (i) *decision nodes*, where a specified player assigns a variable given the values of the parent variables, (ii) *chance nodes*, where nature randomly assigns a variable conditioned on the values of the parent variables (iii) *utility nodes*, where a specified player receives utility as a function of the values of the parent variables. Fig. 1 shows a MAID corresponding to the job market game. It is straightforward to check that MAIDs indeed support all of the queries outlined above (see, for example, the description of how MAIDs represent EFGs [Blum et al., 2006]).

This paper provides a solution to two problems that are not addressed by MAIDs. The first problem is that MAIDs cannot represent context-specific independence. For example, each play in a MAID consists of the same number of actions, whereas in succinct EFGs the number of actions can depend on the actions that have been played. The second problem is that MAID algorithms rely on clique tree representations and thus have both space and time complexity polynomial in the size of the largest clique in this tree, i.e., exponential in treewidth, and therefore superpolynomial in sequence complexity. Our approach employs representations and relies on operations that are strictly polynomial in sequence complexity. We demonstrate these two problems on two examples.

Our first example is a poker variant called Indian poker. In Indian poker, players receive one card each. They hold it against their forehead, so they can see others' cards, but not their own. At the beginning, each player contributes an ante of $1. After the cards are dealt, players take turns of either entering the game and paying an additional $1 (action 'bet') or not entering the game (action 'pass'). If all three players choose 'pass', their ante is returned. If some player chooses 'bet', betting continues until the turn of the player who placed the first bet. Thus, players have an opportunity to match the bet, but they are only allowed to bet once. We consider a version with three players and assume that the deck consists of $C$ cards with distinct values.

A natural representation of Indian poker by a MAID consists of three chance nodes for cards dealt to individual players, five decision nodes for the maximum of five bets, and utility nodes for payoffs to individual players. Since the decision nodes have incoming edges from all the previous bets and from the cards dealt to the other two players, the MAID representation has a sequence complexity $\Gamma_{\text{MAID}} = C^2(2 + 4 + 8 + 16 + 32) = 62C^2$. Using non-

**Input:** succinct EFG, payoff importance vector $\mathbf{w} \in \mathbb{R}^N$
target precision $\varepsilon > 0$, failure probability $\delta > 0$
**Output:** $\varepsilon$-approximate EFCE
Let $\bar{T} := \bar{T}(\varepsilon/3)$, $M_{nat} := M_{nat}(\varepsilon/3, \delta/2)$, $M := M(\varepsilon/3, \delta/2\bar{T})$
$\tilde{p}_{nat} \leftarrow M_{nat}$ independent samples from $p_{nat}$
Let $\boldsymbol{\lambda}_1 = \mathbf{0}$
For $t = 1, 2, \ldots, \bar{T}$:
- $\tilde{q}_t \leftarrow M$ independent samples from $\bar{q}_t(s) \propto e^{\mathbf{w} \cdot \tilde{\mathbf{u}}(s) - \boldsymbol{\lambda}_t \cdot \tilde{\mathbf{r}}(s)}$
- $\phi^* = \operatorname{argmax}_\phi \tilde{\mathbb{E}}_t[\tilde{r}_\phi(s)]$
  $r^* = \tilde{\mathbb{E}}_t[\tilde{r}_{\phi^*}(s)]$
- if $r^* < 2\varepsilon/3$ then return $\tilde{q}_t$
  else $\delta^* = \dfrac{1}{2r_{\max} + 2\varepsilon/3} \ln\left(\dfrac{r_{\max} - r^* + 2\varepsilon/3}{r_{\max} + r^*}\right)$
  $\lambda_{t+1,\phi} = \begin{cases} \lambda_{t,\phi} - \delta^* & \text{if } \phi = \phi^* \\ \lambda_{t,\phi} & \text{otherwise} \end{cases}$

Figure 2: *Main algorithm.*

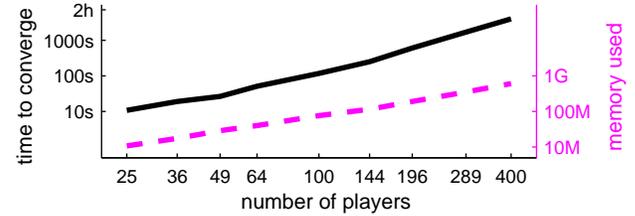

Figure 3: *Grid game experiments.*

MAID succinct EFGs, we can do better: we can terminate the game early if the fourth and fifth bet do not occur. We can also account for the fact that nature must assign different cards to each player. The result is a sequence complexity $\Gamma = 24C(C-1)$. To contrast MAID algorithms with our algorithm further, note that a maximum clique in the MAID representation includes all chance nodes, all decision nodes and exactly one utility node (which takes values from $\{-2, -1, 0, 3, 4\}$), and thus its size is $C^3 \cdot 2^5 \cdot 5 = 160C^3$. For $C = 8$, this equals $8.2 \times 10^4$, whereas the sequence complexity is only $1.3 \times 10^3$.

The second example is the grid game [Vickrey and Koller, 2002], which is played on an $L$-by-$L$ grid by $L^2$ players. Locations on the grid correspond to distinct players. Each player has three actions, and his payoffs depend only on the actions of his neighbors. Specifically, his payoff is the sum of payoffs in independent games with each of his neighbors. Payoffs to each player in each game are specified by a separate 3-by-3 matrix with entries chosen independently uniformly at random from $[0, 1]$. Since the treewith of the resulting graph is $L$, the maximum clique size is $3^L$, whereas the sequence complexity is only $\Gamma = 3L^2$. Our algorithm has space complexity polynomial in $\Gamma$ and therefore achieves significant savings over MAID algorithms.

## 2.2 EQUILIBRIA IN EFGS

Equilibria are solution concepts developed to reason about interactions of utility-maximizing players. They describe situations in which none of the players have incentives to change their behavior. Such stability typically requires randomization.

The most popular equilibrium concept is a *Nash equilibrium*. In a Nash equilibrium, individual players pick their actions at random from a distribution among their best responses to the play of others. For example, in the unique Nash equilibrium of rock-paper-scissors, each player chooses uniformly at random among the three possible actions. Nash equilibria always exist, but many problems related to finding Nash equilibria are NP-hard even for games in strategic form [Conitzer and Sandholm, 2003; Blum et al., 2006]. Moreover, in certain applications, the concept of a Nash equilibrium may be too restrictive because of the assumption that each player randomizes independently. In many real-world situations, players can improve their payoffs by correlating their behavior according to an outside signal; an example is correlation of drivers' behavior at an intersection by a traffic light. Both the computational tractability and correlation among players in one-step games are addressed by a *correlated equilibrium* [Aumann, 1974]. In addition to attractive computational properties, correlated equilibria can be motivated behaviorally as results of adaptive behavior [Hart and Mas-Colell, 2000]. We are not aware of a similar result for Nash equilibria except for trivial cases such as zero-sum games.

Formally, a correlated equilibrium is implemented by an external mechanism called a *moderator* which draws a strategy profile $s$ from some distribution $p(s)$ and reveals the portion $s_n$ to the player $n$. The joint distribution $p(s)$ is called a correlated equilibrium if none of the players can improve their expected utility based on the information provided. Each correlated equilibrium $p(s)$ can also be viewed as a Nash equilibrium in the extended game where nature first draws $s$ and the individual players take actions with the additional information consisting of the suggestion $s_n$.

While finding correlated equilibria is more tractable than finding Nash equilibria, the existing algorithms, such as [Papadimitriou, 2005], were developed for strategic-form games and are therefore polynomial in the number of reduced strategies, which can be exponential in the sequence complexity. Moreover, the mechanism suggested by the correlated equilibrium is unnatural in sequential settings, because the moderator must reveal the entire strategy $s_n$ to each player up front; for example, in a poker game, the moderator would need to tell each player what to do in every possible situation that can arise during the game.

Instead of correlated equilibria we study their generalization to sequential settings called *extensive-form correlated equilibria* [Forges and von Stengel, 2002]. Similar to a correlated equilibrium, an *extensive-form correlated equilibrium* (EFCE) is a probability distribution over strategy profiles implemented by a moderator. The moderator, however, does not reveal entire strategies to players at the beginning of the game. Instead it suggests actions only when players reach the relevant information sets during the game. A player can either follow the moderator's suggestion or deviate. When players deviate, they stop receiving suggestions and must follow their own strategies.

We formalize this behavior by the notion of a *causal deviation* [Gordon et al., 2008]. Causal deviations of the player $n$ are characterized by triples $(i^{trig}, a^{trig}, s_n^{dev})$ as follows: the player $n$ follows suggestions of the moderator until receiving the suggestion $a^{trig}$ in the information set $i^{trig}$. If such a suggestion is received, the player begins to deviate and play according to $s_n^{dev}$. The pair $(i^{trig}, a^{trig})$ is referred to as the *trigger*, the strategy $s_n^{dev}$ as the *deviation strategy*.

For computational and notational convenience, we use an equivalent definition of the EFCE due to Gordon et al. [2008], in which the moderator shows individual players their entire strategies $s_n$, but, distinctly from the correlated equilibrium, players cannot deviate to arbitrary strategies $s'_n$ after seeing $s_n$. Instead they may only apply causal deviations, represented as maps $\phi_{i^{trig}, a^{trig}, s_n^{dev}} : s_n \mapsto s'_n$ such that

$$s'_i = \begin{cases} s_i^{dev} & \text{if } s_{i^{trig}} = a^{trig} \text{ and } i \text{ is reachable from } i^{trig} \\ s_i & \text{otherwise.} \end{cases}$$

Formally, $p$ is an EFCE if and only if for all $n$ and all causal deviations $\phi$ of $n$,

$$\mathbb{E}_p[u_n(s_{-n}, \phi(s_n))] \leq \mathbb{E}_p[u_n(s_{-n}, s_n)] \quad , \tag{1}$$

i.e., deviating from the moderator's suggestion by following the map $\phi$ provides no improvement. (We use $s_{-n}$ to denote the vector of strategies of all agents except $n$.) The pointwise difference

$$u_n(s_{-n}, \phi(s_n)) - u_n(s_{-n}, s_n)$$

is denoted as $r_\phi(s)$ and referred to as the *regret* (of $s$ for not deviating according to $\phi$). The set of causal deviations is denoted $\Phi$ and includes the identity map $Id$, corresponding to no deviation. Eq. (1) thus requires that the expected regret be non-positive for all deviations $\phi \in \Phi$. If the expected regret is bounded above by a positive constant $\varepsilon$, we call the distribution $p$ an $\varepsilon$-*approximate* EFCE or simply $\varepsilon$-EFCE. At this point, we can define a *regret bound* as a value $r_{\max}$ such that $r_{\max} \geq |r_\phi(s)|$ for all $s$ and $\phi$.

An important subtlety in our definition of $\varepsilon$-EFCE is that $\Phi$ only considers a single trigger at a time. By using multiple triggers, a player could gain more. We can bound the total gain of the player $n$ by $\varepsilon \Gamma_n \leq \varepsilon \Gamma$ since the player $n$ has only $\Gamma_n$ triggers. In our experiments below, we set $\varepsilon$ small enough to account for this factor.

We stress that the revelation of the entire strategy and restriction to causal deviations are only for notational and computational convenience. In practice, we still implement an EFCE using a sequential moderator, and players are free to deviate however they wish.

Even though the number of causal deviations $\phi$ is much smaller then the number of all possible deviations, it is still prohibitively large to enumerate explicitly. In principle we need to consider all possible deviation strategies across all triggers, which yields $|\Phi| \approx \sum_n \Gamma_n |S_n^*|$. In next section we show how despite this explosion it is possible to derive representations polynomial in the logarithm of this size, $\log |\Phi| \leq \log(\Gamma |S|) \leq \log \Gamma + \log |S| = O(\Gamma)$.

To understand effects of the distinction between correlated equilibria and EFCEs, consider the job market game. As Forges and von Stengel [2002] show, in this game the only correlated and Nash equilibria are non-cooperative: the student never studies and the employer never hires. On the other hand EFCEs for this game include cooperative outcomes. The problem with correlated equilibria in this game is that the moderator needs to reveal the entire strategy to the student at the beginning of the game and if the student is allowed to choose an arbitrary deviation (not necessarily causal), he will always choose not to study.

## 3 MAIN ALGORITHM

In this section we derive a coordinate descent algorithm for calculating EFCEs. Similar to Nash and correlated equilibria, there may be many distributions satisfying the zero-regret condition of Eq. (1) and hence many EFCEs. Depending on the application, some of these equilibria may be more desirable than others. For example, in mechanism design, we seek equilibria with high social welfare; in game-theoretic modeling, we seek equilibria consistent with observed data. Our algorithm allows exploration of the set of equilibria, which facilitates such tasks.

We calculate equilibria of maximum entropy, with a bias term specifying additional properties we might be interested to optimize. The maximum-entropy condition determines a unique equilibrium and also leads to an efficient coordinate-descent procedure analogous to a popular classification algorithm AdaBoost [Freund and Schapire, 1997, 1999]. Specifically, we consider the optimization problem

$$\max_p \left[ H(p) + \mathbf{w} \cdot \mathbb{E}_p[\mathbf{u}(s)] \right] \tag{2}$$
$$\text{s.t. } \mathbb{E}_p[r_\phi(s)] \leq 0 \text{ for all } \phi \in \Phi$$

where $H(p) = \mathbb{E}_p[-\ln p]$ denotes the entropy, and $\mathbf{w}$ is the vector of importance weights that we attach to the expected utilities of individual players. If $\mathbf{w} = \mathbf{0}$, the unique solution of Eq. (2) is the maximum entropy EFCE. Setting $\mathbf{w} \neq \mathbf{0}$ allows us to explore the set of EFCEs by biasing the solution toward equilibria that maximize a linear combination of the players' utilities. Instead of utilities, we could include expectations of other quantities of interest.

The optimization problem above has $|S|$ variables and $|\Phi|$ constraints. As we argued earlier, these are prohibitively large for even moderately sized games. For example, in Indian poker with eight cards, there are about $10^{168}$ strategy profiles and up to $10^{70}$ causal deviations. Our algorithm manages to solve Eq. (2) with representations only logarithmic in these quantities. We use sampling to address the size of $S$ and dynamic programming to address the size of $\Phi$. Before we describe these approaches, we discuss an idealized version of the algorithm, which assumes that we can enumerate all $s$'s and $\phi$'s.

By Theorem 2 of Dudík et al. [2007], the solution of Eq. (2) is a limit of exponential-family densities of the form

$$q_{\boldsymbol{\lambda}}(s) \propto e^{\mathbf{w}\cdot\mathbf{u}(s)-\boldsymbol{\lambda}\cdot\mathbf{r}(s)}$$

where $\boldsymbol{\lambda} \in [0,\infty)^{\Phi}$ and $\mathbf{r}(s)$ denotes the vector $r_{\Phi}(s)$. In the limit these distributions minimize the partition function $Z_{\boldsymbol{\lambda}} = \sum_s e^{\mathbf{w}\cdot\mathbf{u}(s)-\boldsymbol{\lambda}\cdot\mathbf{r}(s)}$. We solve Eq. (2) by the coordinate-descent algorithm SUMMET [Dudík et al., 2007], which in this particular case reduces to AdaBoost.

AdaBoost (and SUMMET) calculate a sequence of parameters $\boldsymbol{\lambda}_t$ (beginning with $\boldsymbol{\lambda}_1 = \mathbf{0}$) and the corresponding densities $q_t$ parametrized by $\boldsymbol{\lambda}_t$. In the $t$-th iteration (round of boosting), the algorithm finds the deviation $\phi$ with the largest regret on $q_t$ and updates the corresponding coordinate:

$$\lambda_{t+1,\phi} = \lambda_{t,\phi} - \frac{1}{2r_{\max}} \ln\left(\frac{r_{\max} - \mathbb{E}_t[r_\phi(s)]}{r_{\max} + \mathbb{E}_t[r_\phi(s)]}\right) \quad (3)$$

where $\mathbb{E}_t$ is the expectation according to $q_t$.

An attractive feature of this algorithm is that the number of rounds required for the convergence of the expected regret to zero does not depend on the number of deviations. The number of rounds is sensitive to the value of $\mathbf{w}$, but for $\mathbf{w} = \mathbf{0}$ it grows only logarithmically with $|S|$. Specifically, if $p^*$ is the equilibrium solving Eq. (2), we are guaranteed to find $q_{\boldsymbol{\lambda}}$ with regret of less than $\varepsilon$ after at most

$$\frac{D(p^* \parallel q_{\mathbf{0}})}{-\frac{1}{2}\ln(1-\varepsilon^2/r_{\max}^2)} \leq \frac{2r_{\max}^2 D(p^* \parallel q_{\mathbf{0}})}{\varepsilon^2} \equiv T(\varepsilon) \quad (4)$$

rounds, where $D(p \parallel q) = \mathbb{E}_p[\ln(p/q)]$ is the relative entropy (or KL-divergence), and the upper bound follows by the inequality $\ln(1+x) \leq x$. (See [Freund and Schapire, 1999].) For $\mathbf{w} = \mathbf{0}$, the distribution $q_{\mathbf{0}}$ is uniform over $S$ and therefore $D(p^* \parallel q_{\mathbf{0}}) \leq \ln|S|$. Since $\ln|S| \leq \Gamma$, we obtain that $T(\varepsilon)$ grows at most linearly with the sequence complexity. Since in each round only a single coordinate of $\boldsymbol{\lambda}$ is updated, the parameter vector remains sparse.

As mentioned at the outset, the algorithm described so far is impractical because the number of strategy profiles and the number of causal deviations can be very large. To overcome these limitations, we approximate the distribution $q_t$ by a sample from $q_t$ using a Markov-chain Monte Carlo (MCMC) algorithm described in Section 3.3, and search for the best deviation on the sample using the algorithm of Section 3.2. Another problem, which may arise in games with frequent nature moves, is that the regret calculation $\mathbb{E}_{nat}[r(s, s_{nat})]$, which is a subroutine of both the MCMC sampling algorithm and the best deviation search, may be intractable. We address this problem by taking a sample from $p_{nat}$. This is straightforward, because $p_{nat}$ takes a product form. If the size of $I(nat)$ is prohibitively large (it could be easily superpolynomial in $\Gamma$), it is possible to take a more compact sample from a surrogate distribution along the lines of Ng and Jordan [2000].

Let $M$ and $M_{nat}$ denote the number of sampled strategy profiles and scenarios, and $\tilde{q}_t$ and $\tilde{p}_{nat}$ the resulting empirical distributions. Let $\tilde{r}_\phi(s)$ denote the approximate regret calculated using $\tilde{p}_{nat}$. Denote by $M_{nat}(\varepsilon, \delta)$ any value $M_{nat}$ such that with probability at least $1 - \delta$

$$|\tilde{r}_\phi(s) - r_\phi(s)| \leq \varepsilon \text{ for all } \phi, s. \quad (5)$$

Similarly, let $M(\varepsilon, \delta)$ be any value $M$ such that for all distributions $q$ over $S$, if $\tilde{q}$ is an empirical distribution of $M$ independent samples from $q$, then with probability at least $1 - \delta$

$$\left|\mathbb{E}_q[\tilde{r}_\phi(s)] - \mathbb{E}_{\tilde{q}}[\tilde{r}_\phi(s)]\right| \leq \varepsilon \text{ for all } \phi. \quad (6)$$

Balancing the convergence of the original algorithm with approximation errors due to scenario sampling and strategy-profile sampling yields the algorithm in Fig. 2. Next, we describe details of this balancing and prove that our algorithm returns an $\varepsilon$-EFCE with high probability. We use additional notation $\bar{q}_{\boldsymbol{\lambda}}$ for the exponential family based on regrets and utilities calculated using $\tilde{p}_{nat}$ instead of $p_{nat}$.

The first step is a technical claim analogous to Corollary 9 of Freund and Schapire [1999]. The original claim establishes that the algorithm with exact updates (Eq. 3) finds an $\varepsilon$-approximate equilibrium in $T(\varepsilon)$ rounds. Since we cannot evaluate expectations under $q_{\boldsymbol{\lambda}}$ or even under $\bar{q}_{\boldsymbol{\lambda}}$, we work with a lower bound on $\mathbb{E}_{s \sim \bar{q}_t}[\tilde{r}_\phi(s)]$, which we can obtain using the MCMC sample from $\bar{q}_t$. Specifically, we work with updates

$$\lambda_{t+1,\phi} = \lambda_{t,\phi} - \frac{1}{2\tilde{r}_{\max}} \ln\left(\frac{\tilde{r}_{\max} - r_t}{\tilde{r}_{\max} + r_t}\right) \quad (7)$$

where $\tilde{r}_{\max} \geq |\tilde{r}_\phi(s)|$ and $r_t$ is a non-negative lower bound on $\mathbb{E}_{\bar{q}_t}[\tilde{r}_\phi]$, i.e., $0 \leq r_t \leq \mathbb{E}_{\bar{q}_t}[\tilde{r}_\phi]$. In our algorithm, we pick $\phi$ for which we can find the largest lower bound $r_t$.

**Claim 1.** *Let $\bar{p}^*$ be the solution of Eq. (2) with $r_\phi$ replaced by $\tilde{r}_\phi$. If the update of Eq. (7) is used in each round, the number of iterations in which $r_t \geq \varepsilon$ is at most*

$$\frac{D(\bar{p}^* \parallel \bar{q}_{\mathbf{0}})}{-\frac{1}{2}\ln(1-\varepsilon^2/\tilde{r}_{\max}^2)} \leq \frac{2\tilde{r}_{\max}^2 D(\bar{p}^* \parallel \bar{q}_{\mathbf{0}})}{\varepsilon^2} \equiv \bar{T}(\varepsilon) \ . \quad (8)$$

Claim 1 follows by a minor modification of the proof of Corollary 9 of Freund and Schapire [1999] and its proof is therefore omitted. Similar to Eq. (4), for $\mathbf{w} = \mathbf{0}$, we obtain that $\bar{T}(\varepsilon) = O(\Gamma/\varepsilon^2)$. We can now state the theorem about convergence of our algorithm.

**Theorem 2.** *Let functions $M(\varepsilon, \delta)$, $M_{nat}(\varepsilon, \delta)$, and $\bar{T}(\varepsilon)$ be defined as above. Then the algorithm of Fig. 2 returns an $\varepsilon$-approximate EFCE with probability at least $1 - \delta$.*

*Proof.* We only analyze the case when Eq. (5) holds for $\tilde{p}_{nat}$ with $\varepsilon$ replaced by $\varepsilon/3$, and Eq. (6) holds for all $\tilde{q}_t$ with $q$ replaced by $\bar{q}_t$ and $\varepsilon$ by $\varepsilon/3$. By the union bound and definitions of $M_{nat}$ and $M$, this happens with probability at least $1 - \delta$. In this case, $\tilde{r}_{\max}$ can be set to $r_{\max} + \varepsilon/3$. Whenever the algorithm performs an update, we have by Eq. (6) that $|\mathbb{E}_{\bar{q}_t}[\tilde{r}_{\phi^*}(s)] - r^*| \le \varepsilon/3$. Since $r^* \ge 2\varepsilon/3$, we obtain that $\varepsilon/3 \le r^* - \varepsilon/3 \le \mathbb{E}_{\bar{q}_t}[\tilde{r}_{\phi^*}(s)]$. By Claim 1 this can happen at most $\bar{T}(\varepsilon/3)$ times. Thus, the algorithm encounters $r^* < 2\varepsilon/3$ during its $\bar{T}(\varepsilon/3)$ rounds and returns the corresponding $\tilde{q}_t$. Since $\tilde{\mathbb{E}}_t[\tilde{r}_\phi(s)] < 2\varepsilon/3$ for all $\phi$, by Eq. (5) we also obtain that $\tilde{\mathbb{E}}_t[r_\phi(s)] \le \varepsilon$. Thus, the returned distribution $\tilde{q}_t$ is an $\varepsilon$-approximate EFCE. □

In practice, if the algorithm does not find an $\varepsilon$-EFCE in $\bar{T}$ rounds, we do not terminate, but continue until the approximate regret drops below a specified value. To complete the algorithm, we need to provide settings for $M$, $M_{nat}$, and implement MCMC sampling and the best deviation search. By Hoeffding's inequality and the union bound, we can use

$$M_{nat}(\varepsilon, \delta) = \frac{2r_{\max}^2 \ln(|S||\Phi|/\delta)}{\varepsilon^2}, M(\varepsilon, \delta) = \frac{2\tilde{r}_{\max}^2 \ln(|\Phi|/\delta)}{\varepsilon^2}.$$

To derive $M(\varepsilon, \delta)$, we can assume that $\tilde{r}_{\max} = r_{\max} + \varepsilon'$ for a suitable $\varepsilon'$ (as we did in the proof of Theorem 2). From the previous section, $\ln|S| \le \Gamma$ and $\ln|\Phi| = O(\Gamma)$. Hence we obtain that both $M$ and $M_{nat}$ scale at most linearly with sequence complexity. The two remaining pieces of our algorithm are the best deviation search and MCMC sampling, which we describe next.

### 3.1 SIMULATION AND BEST-RESPONSE TREES

Both our best-deviation calculation and our MCMC algorithm rely on tree representations derived from the game tree restricted to a sample of strategy profiles and a sample of scenarios. Let $\mathcal{T}$ denote the game tree. For a set $\tilde{S}$ of strategy profiles and a set $\tilde{S}_{nat}$ of scenarios, we define the *simulation tree* $\mathcal{T}(\tilde{S}, \tilde{S}_{nat})$ as the tree of possible plays when players and nature are restricted to play only according to strategy profiles and scenarios in $\tilde{S}$ and $\tilde{S}_{nat}$. If strategy profiles and scenarios are drawn from distributions supported on $\tilde{S}$ and $\tilde{S}_{nat}$ we say that players and nature are *simulated* by the respective distributions. We write $\mathcal{T}^h(\tilde{S}, \tilde{S}_{nat})$ for the subtree of the simulation tree beginning in the node $h$.

In addition to simulation trees, we also consider *best-response trees* representing *best-response* games, in which all players except for one are simulated by some distribution. The unsimulated player maximizes his utility by playing the best response to the simulated players. For notational convenience, we define best-response trees relative to full strategy profiles. For a single strategy profile $s$ and a single scenario $s_{nat}$, the best-response tree $\mathcal{T}_n(s, s_{nat})$ is the union of paths $\mathcal{T}((s'_n, s_{-n}), s_{nat})$ across all $s'_n \in S_n$. The best-response tree for $\tilde{S}$ and $\tilde{S}_{nat}$, written as $\mathcal{T}_n(\tilde{S}, \tilde{S}_{nat})$, is the union of $\mathcal{T}_n(s, s_{nat})$ across all $s \in \tilde{S}$ and $s_{nat} \in \tilde{S}_{nat}$. Similar to simulation trees, we also consider best-response subtrees beginning at a concrete node $h$.

The *best-response complexity* for the player $n$ is defined as the maximum size of $\mathcal{T}_n(s, s_{nat})$

$$\Lambda_n = \sup_{s, s_{nat}} |\mathcal{T}_n(s, s_{nat})| \ .$$

The best-response complexity of a game is defined as $\Lambda = \max_n \Lambda_n$. If $d$ denotes the depth of the game tree, we obtain that $\Lambda_n/d \le \Gamma_n$. Note that $\Lambda_n$ is approximately exponential in the number of actions that the player $n$ can take during the game, whereas $\Gamma_n$ is also exponential in the amount of information that the player can observe. Thus, it is straightforward to construct examples when $\Lambda_n \ll \Gamma_n$.

### 3.2 BEST DEVIATION

Recall that EFCE is characterized by the zero-regret condition over causal deviations and each causal deviation $\phi$ is described by a trigger $(i^{trig}, a^{trig})$ and a deviation strategy $s_n^{dev}$. The key challenge is that the number of deviation strategies can be very large. We overcome this by dynamic programming.

We extend the simulation tree for a given set of scenarios and strategy profiles, so that when simulating an action $a$ in an information set $i$, the simulated player is allowed to deviate from simulation and enter the best-response tree beginning in that node. We refer to the resulting tree as the *deviation tree* and the corresponding game as the *deviation game*. The best response of each player in the deviation game consists of a set of triggers (where the player benefits by deviating) and, for each trigger, the best deviation strategy. It is straightforward to verify that thanks to the perfect recall assumption, the best deviation and the corresponding regret can be calculated by dynamic programming in time linear in the size of the deviation tree. To derive an upper bound on this size, first consider a single scenario $s_{nat}$ and a single strategy profile $s$. The resulting deviation tree consists of at most $d$ simulated actions, each possibly giving rise to a best-response tree of size at most $\Lambda$. Summing across all possible scenario-profile pairs we find that the deviation tree size is at most $O(d\Lambda MM_{nat})$, which is polynomial in $\Gamma$.

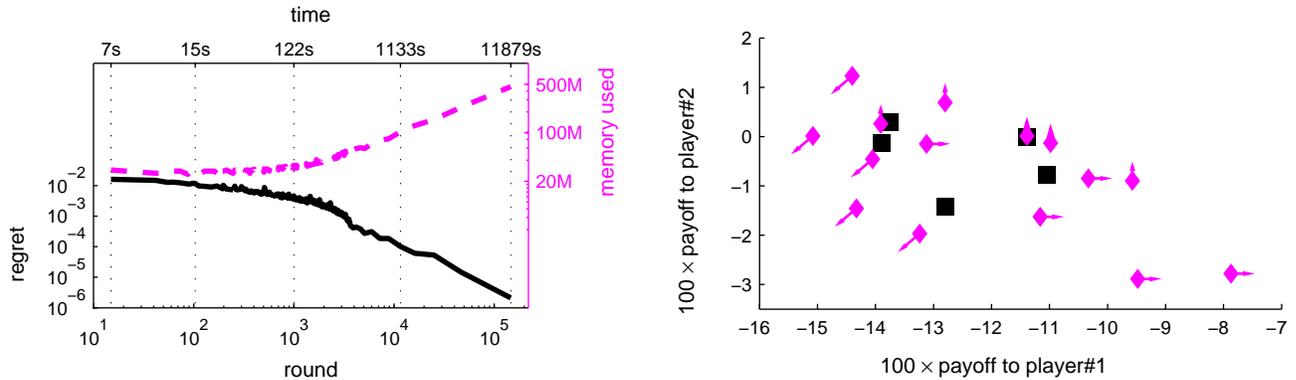

Figure 4: *Indian poker experiments. Left:* Log-log plot of regret and memory usage with an increasing number of rounds. After the initial period of about 1000 rounds, the regret $r^*$ appears to converge much faster than predicted by the bound $r^* \propto 1/\sqrt{t}$. *Right:* Equilibria payoffs for various settings of the weight vector, for multiple randomization seeds of the MCMC algorithm. Squares indicates the setting of $\mathbf{w} = \mathbf{0}$ (maximum entropy runs); diamonds indicate the remaining runs with arrows pointing in the direction of the weight vector $\mathbf{w}$. As expected, the equilibrium payoffs are approximately biased in the direction of $\mathbf{w}$ relative to the maximum equilibrium solutions.

### 3.3 MCMC SAMPLING

The goal of the MCMC sampling step is to obtain a sample from the current EFCE estimate $\bar{q}_t$. In our sampling algorithm, we do not work with full strategy profiles $s$, but instead work with their partial versions which we call *skeletons*. They are the minimum portions of strategy profiles needed to evaluate $\bar{q}_t(s)$. More precisely, the skeleton of a strategy profile $s$ relative to a set of scenarios $\tilde{S}_{nat}$ and a set of causal deviations $\Psi$ with $Id \in \Psi$ is defined as the partial strategy profile $s_{skel}$ such that $(s_{skel})_i = s_i$ if the information set $i$ is reachable under some combination of a deviation and a scenario from $\Psi$ and $\tilde{S}_{nat}$, and $(s_{skel})_i = *$ otherwise. Thus, $\tilde{\mathbf{u}}(s_{skel})$ and $\tilde{r}_\Psi(s_{skel})$ are well defined and equal to $\tilde{\mathbf{u}}(s)$ and $\tilde{r}_\Psi(s)$. We consider skeletons relative to the set of scenarios sampled at the beginning of the algorithm and the set of deviations $\Psi$ containing the identity and the best deviations from the first $t$ rounds of the algorithm.

Formally, we sample full strategy profiles using slice Metropolis-Hastings (MH) sampling (see, for example, [Andrieu et al., 2003]). Slices in our case correspond to sets of strategy profiles $s$ with identical skeletons. We alternatingly transition from a strategy profile $s$ to $s'$ according to the MH algorithm, and from $s'$ to $s''$ uniformly within the same slice. We use slice sampling for two reasons. On one hand, it improves mixing of the Markov chain. On the other hand, it means that when transitioning from $s'$ to $s''$, we can "forget" actions in all information sets except for those in the skeleton. When accessing non-skeleton information sets in the next MH transition, the corresponding action $s_i$ is sampled uniformly from $\{1, \ldots, A_i\}$.

To implement MH sampling, we need to define a proposal distribution $\hat{q}(s' \mid s)$. We implement $\hat{q}$ by first randomly picking one of the information sets in $s_{skel}$, and then flipping the action $s_i$ according to target probabilities $\bar{q}_t(s')$ where $s' = (s_{-i}, s'_i)$ with $s'_i \in \{1, \ldots, A_i\}$. Formally, we assume a predefined conditional probability $\hat{q}(i \mid s)$, zero whenever $(s_{skel})_i = *$, and define

$$\hat{q}(s' \mid s) = \sum_i \hat{q}(i \mid s) \mathbb{1}(s'_{-i} = s_{-i}) \frac{\bar{q}_t(s_{-i}, s'_i)}{\sum_{s''_i} \bar{q}_t(s_{-i}, s''_i)}$$

$$= \sum_i \hat{q}(i \mid s) \mathbb{1}(s'_{-i} = s_{-i}) \frac{e^{\mathbf{w} \cdot \tilde{\mathbf{u}}(s_{-i}, s'_i) - \boldsymbol{\lambda}_t \cdot \tilde{\mathbf{r}}(s_{-i}, s'_i)}}{\sum_{s''_i} e^{\mathbf{w} \cdot \tilde{\mathbf{u}}(s_{-i}, s''_i) - \boldsymbol{\lambda}_t \cdot \tilde{\mathbf{r}}(s_{-i}, s''_i)}}.$$

We use $\hat{q}(i \mid s)$ uniform over $i$ in the skeleton of $s$. The resulting acceptance probability is $\min\bigl(1, |s_{skel}|/|s'_{skel}|\bigr)$ where $|s_{skel}|$ and $|s'_{skel}|$ denote the sizes of supports of the given skeletons (the numbers of entries different from $*$).

For an efficient implementation of sampling, we use a simulation tree $\mathcal{T}(s, \tilde{S}_{nat})$ where $s$ is the random variable of the Markov chain. We calculate regret values $\tilde{r}_\phi$ for $\phi_{i^{trig}, a^{trig}, s_n^{dev}} \in \Psi$ using simulation subtrees $\mathcal{T}^h(s, \tilde{S}_{nat})$ as well as subtrees corresponding to the deviation $\mathcal{T}^h\bigl((s_{-n}, s_n^{dev}), \tilde{S}_{nat}\bigr)$, where $h$ is taken from the set of nodes corresponding to the suggestion $a^{trig}$ in $i^{trig}$. As before, we organize computations for all $\phi \in \Psi$ in a single tree whose size is at most $O(d\Lambda M_{nat})$. While building this tree requires time linear in its size, steps of the Markov chain affect only small portions of it. Again, space complexity is at most polynomial in $\Gamma$.

### 3.4 IMPROVING CONVERGENCE

Despite favorable convergence guarantees, in initial versions of our experiments (Section 4) we found that the algorithm of Fig. 2 may be too slow in practice. Instead of the update suggested by the algorithm, we perform a line search to find the update that decreases the dual objective $Z_\lambda$ the most (the normalization is only over the support of $\tilde{q}_t$ and therefore is not prohibitively expensive). Empirically, this leads to large improvements in the rate of convergence (see Section 4). While this strategy does not break the convergence proof for the exact-update algorithm, in approximate settings it may overfit the current sample $\tilde{q}_t$. Therefore, we allow at most a constant-fold increase over the original update and also stop the line search at the point when the "effective sample size" of the reweighted distribution would drop below a specified threshold (see below).

Another source of inefficiency is MCMC sampling at the beginning of each round. The sampling step is computationally expensive and potentially quite wasteful, because in each round the distribution $\bar{q}_t$ changes only slightly. To avoid wasteful sampling, we use a reweighted $\tilde{q}_t$ for several iterations until the logarithm of the effective sample size (measured by the entropy of the weights) drops below a specified threshold (similar to Welling et al. [2003] and Broderick et al. [2007]). Further substantial speedup comes from employing the same deviation tree between MCMC sampling steps. If carefully implemented, the intermediate results of dynamic programming can be reused in consecutive rounds. The combination of the MCMC-sample reuse and efficient data structures leads to a running-time decrease by several orders of magnitude. (Details about the efficient implementation of weight updates will be provided in the extended version.)

The final modification to the basic algorithm comes from the observation that in the initial rounds it is wasteful to use a large number of samples $M$. In fact the number of sampled strategy profiles should be just sufficient to determine the deviation with the largest regret. To determine regret to the precision of $\varepsilon \approx r^*$ we need no more than $O(1/\varepsilon^2) = O(1/r^{*2})$ samples. From Claim 1, we expect that $r^* = O(1/\sqrt{t})$. Hence, it suffices to increase $M$ linearly with $t$. In our experiments, we use sample size $100 + t/10$.

## 4 EXPERIMENTS

We evaluate our algorithm on three games introduced earlier: the job market game, an eight-card version of Indian poker, and the grid game. In Indian poker, which is the only game with chance moves, we enumerate all possible scenarios (the total of 336 deals) and thus obtain exact regrets. All our experiments were performed on an Intel Core 2 Duo processor running at 2.66GHz with 4GB of RAM.

We ran our algorithm on the job market game with weight vectors $\mathbf{w} = (0, 0)$ and $\mathbf{w} = (1, 1)$, ten times for each, with different random seeds for the MCMC algorithm. The former case corresponds to maximizing entropy, the latter to maximizing the sum of entropy and social welfare (the sum of payoffs). In all cases our algorithm converged to an exact equilibrium in less than 15 rounds. In the maximum entropy case, we observed some variance among solutions.

In ten runs, the payoff to the student ranged from 2.1 to 2.7, the payoff to the employer from 0.1 to 0.5. In each equilibrium we observed significant correlation between the student and the employer. The probability of cooperative behavior (the student studies and then gets hired) ranged between 26% and 36%. Results in the second case were qualitatively different. The algorithm always found a fully cooperative solution with payoffs $(4, 5)$ where the student always studies and always gets hired. The equilibrium distribution itself, however, varied in relative proportions of the correct answer being 'yes' or 'no'.

Indian poker is more challenging: unlike the job market game with four reduced strategies for each player and the total of 16 reduced strategy profiles, Indian poker admits $9^{C(C-1)}$, $10^{C(C-1)}$, and $16^{C(C-1)}$ reduced strategies, respectively, for its three players, yielding the total of more than $10^{3C(C-1)}$ reduced strategy profiles. Thus, for $C = 8$, each player has at least $10^{53}$ reduced strategies and the total number of reduced strategy profiles is more than $10^{168}$. Obviously, strategic-form algorithms cannot be applied in this case. On the other hand, the sequence complexity for $C = 8$ is only $1.3 \times 10^3$, so our algorithm can solve this problem easily.

We ran our algorithm for the maximum entropy case as well as for weight vectors $(100, 0, 0)$, $(0, 100, 0)$, and $(0, 0, 100)$, each for five different MCMC random seeds. We explored the convergence of regret, space complexity, and the type of solutions. In Fig. 4, on the left, we show log-log plots reporting convergence of the regret and an increase in memory use as a function of the number of rounds. We stopped the algorithm when regret dropped to about $10^{-4}$ of its initial value. According to the bound of Eq. (4), log regret should be decreasing with the slope $-0.5$. However, after initial 1000 rounds, the convergence dramatically improves and the slope appears to be around $-1.5$; i.e., $r^* \propto t^{-1.5}$ rather than $r^* \propto t^{-0.5}$.

On the right-hand side of Fig. 4, we plot the resulting payoff vectors. In all cases, *player1* on average loses money to *player2* and *player3*. This is to be expected since the game is asymmetric and players that move later in the game have more information. Arrows in the plot correspond to nonzero weight vectors; since the sum of payoffs is 0, weight vectors $(w_1, w_2, w_3)$ and $(w_1 - w_3, w_2 - w_3, 0)$ are equivalent; we plot the latter. As expected, for nonzero vectors $\mathbf{w}$, equilibrium payoffs are pushed from the maximum entropy solution approximately in the direction of $\mathbf{w}$.

Finally, we explore the performance of our algorithm on grid games with sizes ranging from 5-by-5 to 20-by-20. These games are particularly challenging for exact approaches, such as the MAID algorithm of Blum et al. [2006], which are polynomial in the size of the largest clique. For our largest grid, the clique size is $3^{20} \approx 3.5 \times 10^9$, which is infeasible for exact approaches. For each grid size, we randomly generated five games and ran our algorithm to the convergence of $10^{-3}$ of the initial regret. In Fig. 3, we report average running times and memory use (standard deviations are approximately equal to the thickness of the line). Log-log plots indicates polynomial dependence on $\Gamma$ of both the running time (slightly faster than quadratic) and the memory use (faster than linear but slower than quadratic). For a comparison, note that Blum et al. [2006] report the performance of an exact algorithm only up to the size 6-by-6, corresponding to the clique size 729 (with the reported running time 80 s on a single-core Intel Xeon).

## 5 CONCLUSION

We have introduced an equilibrium calculation approach for sequential games that is based on coordinate descent and MCMC sampling. Unlike previous approaches, our algorithm relies on representations polynomial in sequence complexity. There are many possible directions for future research, such as incorporating more sophisticated best-deviation algorithms or improving sampling. Another alternative is to consider smaller sets of deviations, or possibly to formulate the best-deviation problem in a way that would allow easy application of external algorithms, akin to base learners in boosting. Many bounds in our analysis, such as bounds on the number of sampled scenarios and strategy profiles, are quite loose and a tighter analysis might provide better understanding of parameters contributing to complexity of equilibrium calculation. For example, it would be interesting to outline cases when our algorithm (or its modification) is polynomial in the best-response complexity $\Lambda$ rather than the sequence complexity $\Gamma$. Finally, it would be of great interest to explore alternatives to sampling, such as variational approximations, belief propagation, or lifted inference.

### Acknowledgements

The authors gratefully acknowledge support from the DARPA grant number HR0011-07-10026, the Computer Science Study Panel program, and the ARO grant number W911NF-08-1-0301.